\documentclass[12pt]{article}
\usepackage{graphicx}
\usepackage{epsfig}
\usepackage{epstopdf}
\DeclareGraphicsExtensions{.pdf,.eps,.png,.jpg,.mps}

\def\lsim{\mathrel{\rlap{\lower3pt\hbox{\hskip0pt$\sim$}}
     \raise1pt\hbox{$<$}}}         
\def\gsim{\mathrel{\rlap{\lower4pt\hbox{\hskip1pt$\sim$}}
     \raise1pt\hbox{$>$}}}         

\usepackage{amsmath}
\usepackage{amsfonts}

\begin{document}
\begin{titlepage}

\centerline{\Large \bf Combining Alphas via Bounded Regression}
\medskip

\centerline{Zura Kakushadze$^\S$$^\dag$\footnote{\, Zura Kakushadze, Ph.D., is the President of Quantigic$^\circledR$ Solutions LLC,
and a Full Professor at Free University of Tbilisi. Email: \tt zura@quantigic.com}}
\bigskip

\centerline{\em $^\S$ Quantigic$^\circledR$ Solutions LLC}
\centerline{\em 1127 High Ridge Road \#135, Stamford, CT 06905\,\,\footnote{\, DISCLAIMER: This address is used by the corresponding author for no
purpose other than to indicate his professional affiliation as is customary in
publications. In particular, the contents of this paper
are not intended as an investment, legal, tax or any other such advice,
and in no way represent views of Quantigic® Solutions LLC,
the website \underline{www.quantigic.com} or any of their other affiliates.
}}
\centerline{\em $^\dag$ Free University of Tbilisi, Business School \& School of Physics}
\centerline{\em 240, David Agmashenebeli Alley, Tbilisi, 0159, Georgia}
\medskip
\centerline{(January 7, 2015; revised October 22, 2015)}

\bigskip
\medskip

\begin{abstract}
{}We give an explicit algorithm and source code for combining alpha streams via bounded regression. In practical applications typically there is insufficient history to compute a sample covariance matrix (SCM) for a large number of alphas. To compute alpha allocation weights, one then resorts to (weighted) regression over SCM principal components. Regression often produces alpha weights with insufficient diversification and/or skewed distribution against, {\em e.g.}, turnover. This can be rectified by imposing bounds on alpha weights within the regression procedure. Bounded regression can also be applied to stock and other asset portfolio construction. We discuss illustrative examples.

\end{abstract}

\bigskip
\medskip

\noindent{}{\bf Keywords:} hedge fund, alpha stream, alpha weights, portfolio turnover, investment allocation, weighted regression, diversification, bounds, optimization, factor models

\end{titlepage}

\newpage

\section{Introduction}

{}With technological advances there is an ever increasing number of alpha streams.\footnote{\, Here ``alpha'' -- following the common trader lingo -- generally means any reasonable ``expected return'' that one may wish to trade on and is not necessarily the same as the ``academic'' alpha. In practice, often the detailed information about how alphas are constructed may not be available, e.g., the only data available could be the position data, so ``alpha'' then is a set of instructions to achieve certain stock holdings by some times  $t_1,t_2,\dots$} Many of these alphas are ephemeral, with relatively short lifespans. As a result, in practical applications typically there is insufficient history to compute a sample covariance matrix (SCM) for a large number of alpha streams -- SCM is singular. Therefore, directly using SCM in, say, alpha portfolio optimization is not a option.\footnote{\, For a partial list of hedge fund literature, see, {\em e.g.}, \cite{HF1}-\cite{HF20} and references therein. For a partial list of portfolio optimization and related literature, see, {\em e.g.}, \cite{PO1}-\cite{PO35} and references therein.}

{}One approach to circumvent this difficulty is to build a factor model for alpha streams \cite{AlphaFM}. Because of the utmost secrecy in the alpha business, such factor models must be build in-house -- there are no commercial providers of ``standardized" factor models for alpha streams. As with factor models for equities, such model building for alphas requires certain nontrivial expertise and time expenditure.

{}Therefore, in practice, one often takes a simpler path. As was discussed in more detail in \cite{AlphaFM}, one can deform SCM such that it is nonsingular, and then use the so-deformed SCM in, say, Sharpe ratio optimization for a portfolio of alphas. For small deformations this then reduces to a cross-sectional weighted regression of the alpha stream expected returns \cite{AlphaFM}. The regression weights are the inverse sample variances of the alphas. The columns of the loadings matrix, over which the expected returns are regressed, are nothing but the first $K$ principal components of SCM corresponding to its positive ({\em i.e.}, non-vanishing) eigenvalues \cite{AlphaFM}.

{}Regression often produces alpha weights with insufficient diversification and/or skewed distribution against, {\em e.g.}, turnover. Thus, if some expected returns are skewed, then, despite nonunit regression weights (which suppress more volatile alphas), the corresponding alpha weights can be larger than desired by diversification considerations. Also, the principal components know nothing about quantities such as turnover.\footnote{\, One approach to rectify this is to add a turnover-based factor to the loadings matrix \cite{AlphaFM}.} A simple way of obtaining a more ``well-rounded" portfolio composition is to set bounds on alpha weights. This is the approach we discuss here.

{}When individual alpha streams are traded on separate execution platforms, the alpha weights are non-negative. By combining and trading multiple alpha streams on the same execution platform -- the framework we adapt here -- one saves on transaction costs by internally crossing trades between different alpha streams (as opposed to going to the market).\footnote{\, For a recent discussion, see \cite{OD}.} Then the alpha weights can be negative.

{}When alpha weights can take both positive and negative values, the bounded regression problem simplifies. It boils down to an iterative algorithm we discuss in Section \ref{sec2}. This algorithm can actually be derived from an optimization algorithm with bounds (in a factor model context) discussed in \cite{MeanRev} by taking the regression limit of optimization. We also give R source code for the bounded regression algorithm in Appendix \ref{appA}. Appendix \ref{appB} contains some legalese. We conclude in Section \ref{sec3}, where we also discuss bounded regression with transaction costs following \cite{AlphaOpt}.

\section{Bounded Regression}\label{sec2}
\subsection{Notations}

{}We have $N$ alphas $\alpha_i$, $i=1,\dots,N$. Each alpha is actually a time series $\alpha_i(t_s)$, $s=0,1,\dots,M$, where $t_0$ is the most recent time. Below $\alpha_i$ refers to $\alpha_i(t_0)$.

{}Let $C_{ij}$ be the sample covariance matrix (SCM) of the $N$ time series $\alpha_i(t_s)$. If $M < N$, then only $M$ eigenvalues of $C_{ij}$ are non-zero, while the remainder have ``small" values, which are zeros distorted by computational rounding.\footnote{\, Actually, this assumes that there are no N/As in any of the alpha time series. If some or all alpha time series contain N/As in non-uniform manner and the correlation matrix is computed by omitting such pair-wise N/As, then the resulting correlation matrix may have negative eigenvalues that are not zeros distorted by computational rounding.}

{}Alphas $\alpha_i$ are combined with weights $w_i$. Any leverage is included in the definition of $\alpha_i$, {\em i.e.}, if a given alpha labeled by $j\in\{1,\dots,N\}$ before leverage is $\alpha^\prime_j$ (this is a raw, unlevered alpha) and the corresponding leverage is $L_j:1$, then we define $\alpha_j\equiv L_j~\alpha^\prime_j$. With this definition, the weights satisfy the condition
\begin{equation}\label{w.norm}
 \sum_{i=1}^N \left|w_i\right| = 1
\end{equation}
Here we allow the weights to be negative as we are interested in the case where the alphas are traded on the same execution platform and trades between alphas are crossed, so one is actually trading the combined alpha $\alpha\equiv\sum_{i=1}^N \alpha_i~w_i$.

\subsection{Weighted Regression}

{}When SCM $C_{ij}$ is singular and no other matrix ({\em e.g.}, a factor model) to replace it is available, one can deform SCM such that it is nonsingular, and then use the so-deformed SCM in, say, Sharpe ratio optimization for a portfolio of alphas \cite{AlphaFM}. For small deformations this reduces to a cross-sectional weighted regression of the alpha stream expected returns \cite{AlphaFM}. The regression weights $z_i$ (not to be confused with the alpha weights $w_i$) are the inverse sample variances of the alphas: $z_i \equiv 1/C_{ii}$. The columns of the loadings matrix $\Lambda_{iA}$, $A=1,\dots,K$, over which the expected returns are regressed, are nothing but the first $K$ principal components of SCM corresponding to its positive ({\em i.e.}, non-vanishing) eigenvalues. However, for now we will keep $\Lambda_{iA}$ general ({\em e.g.}, one may wish to include other risk factors in $\Lambda_{iA}$ \cite{AlphaFM}).

{}The weights $w_i$ are given by:
\begin{equation}\label{w-reg}
 w_i = \gamma~z_i~\varepsilon_i
\end{equation}
where $\varepsilon_i$ are the residuals of the cross-sectional regression of $\alpha_i$ over $\Lambda_{iA}$ (without the intercept, unless the intercept is subsumed in $\Lambda_{iA}$, that is -- see below) with the regression weights $z_i$:
\begin{equation}\label{residuals}
 \varepsilon_i = \alpha_i - \sum_{j = 1}^N z_j~\alpha_j \sum_{A,B = 1}^K \Lambda_{iA}~\Lambda_{jB}~Q^{-1}_{AB}
\end{equation}
where $Q^{-1}_{AB}$ is the inverse of
\begin{equation}
 Q_{AB}\equiv \sum_{i = 1}^N z_i~\Lambda_{iA}~\Lambda_{iB}
\end{equation}
and the overall factor $\gamma$ in (\ref{w-reg}) is fixed via (\ref{w.norm}). Note that we have
\begin{equation}\label{lin.const}
 \forall A\in\{1,\dots,K\}:~~~\sum_{i=1}^N w_i~\Lambda_{iA} = 0
\end{equation}
So, the weights $w_i$ are neutral w.r.t. the risk factors defined by the columns of the loadings matrix $\Lambda_{iA}$.

\subsection{Bounds}

{}Since the weights $w_i$ can have either sign, we will assume that the lower and upper bounds on the weights
\begin{equation}
 w^-_i \leq w_i \leq w^+_i
\end{equation}
satisfy the conditions
\begin{eqnarray}
 &&w^-_i \leq 0\\
 &&w^+_i \geq 0\\
 &&w^-_i < w^+_i
\end{eqnarray}
The last condition is not restrictive: if for some alpha labeled by $i$ we have $w^-_i = w^+_i$, then we can simply set $w_i = w^-_i$ and altogether exclude this alpha from the bounded regression procedure below. Also, if, for whatever reason, we wish to have no upper/lower bound for a given $w_i$, we can simply set $w^\pm_i = \pm 1$.

{}The bounds can be imposed for diversification purposes: {\em e.g.}, one may wish to require that no alpha has a weight greater than some fixed (small) percentile $\xi$, {\em i.e.}, $|w_i|\leq \xi$, so $w^\pm_i = \pm \xi$. One may also wish to suppress the contributions of high turnover alphas, {\em e.g.}, by requiring that $|w_i| \leq {\widetilde \xi}$ if $\tau_i\geq \tau_*$, where $\tau_i$ is the turnover,\footnote{\, Here the turnover (over a given period, {\em e.g.}, daily turnover) is defined as the ratio $\tau_i \equiv D_i/I_i$ of total dollars $D_i$ (long plus short) traded by the alpha labeled by $i$ over the corresponding total dollar holdings $I_i$ (long plus short).} $\tau_*$ is some cut-off turnover, and ${\widetilde\xi}$ is some (small) percentile. Bounds can also be used to limit the weights of low capacity\footnote{\, By capacity $I^*_i$ for a given alpha we mean the value of the investment level $I_i$ for which the P\&L $P_i(I_i)$ is maximized (considering nonlinear effects of impact).} alphas. {\em Etc.}\footnote{\, Since the regression we consider here is weighted with the regression weights $z_i=1/C_{ii}$, this already controls exposure to alpha volatility, so imposing bounds based on volatility would make a difference only if one wishes to further suppress volatile alphas.}

\subsection{Running a Bounded Regression}

{}So, how do we impose the bounds in the context of a regression? There are two subtleties here. First, we wish to preserve the factor neutrality property (\ref{lin.const}), which is invariant under the simultaneous rescalings $w_i\rightarrow \zeta w_i$ (where $\zeta$ is a constant). If we simply set some $w_i$ to their upper or lower bounds, this generally will ruin the rescaling invariance, so the property (\ref{lin.const}) will be lost. Second, we must preserve the normalization condition (\ref{w.norm}). In fact, it is precisely this normalization condition that allows to meaningfully set the bounds $w^\pm_i$, as the regression itself does not fix the overall normalization coefficient $\gamma$ in (\ref{w-reg}), owing to the rescaling invariance $w_i\rightarrow \zeta w_i$.

{}Here we discuss the bounded regression algorithm. To save space, we skip the detailed derivation as it follows straightforwardly by taking the regression limit of optimization with bounds in the context of a factor model, both of which are discussed in detail in \cite{MeanRev}.\footnote{\, The regression limit of optimization essentially amounts to the limit $\xi^2_i \equiv \eta~{\widetilde \xi}^2_i$, $\eta\rightarrow 0$, ${\widetilde \xi}^2_i = \mbox{fixed}$, where $\xi_i$ is the specific (idiosyncratic) risk in the factor model with the factor loadings matrix identified with the regression loadings matrix $\Lambda_{iA}$ (and the $K\times K$ factor covariance matrix becomes immaterial in the regression limit) -- see \cite{MeanRev} for details.}

{}Let us define the following subsets of the index $i\in J\equiv \{1,\dots,N\}$:
\begin{eqnarray}
 &&w_i = w_i^+,~~~i\in J^+\\
 &&w_i = w_i^-,~~~i\in J^-\\
 &&{\overline J} \equiv J^+ \cup J^-\\
 &&{\widetilde J} \equiv J \setminus {\overline J}
\end{eqnarray}
Further, let
\begin{eqnarray}
 && {\widetilde \alpha}_i \equiv \gamma~\alpha_i\\
 && y_A \equiv \sum_{i\in{\widetilde J}} z_i~{\widetilde \alpha}_i~\Lambda_{iA} + \sum_{i\in J^+} w_i^+~\Lambda_{iA} + \sum_{i\in J^-} w_i^-~\Lambda_{iA}\label{y}
\end{eqnarray}
where $\gamma$ is to be determined (see below). Then we have
\begin{eqnarray}\label{wv.bounds}
 && w_i = z_i \left({\widetilde \alpha}_i - \sum_{A,B=1}^K \Lambda_{iA}~{\widetilde Q}^{-1}_{AB}~y_B\right),~~~i\in {\widetilde J}\\
 && \forall i\in J^+:~~~z_i \left({\widetilde \alpha}_i - \sum_{A,B=1}^K \Lambda_{iA}~{\widetilde Q}^{-1}_{AB}~y_B\right) \geq w_i^+\label{Jp}\\
 && \forall i\in J^-:~~~z_i \left({\widetilde \alpha}_i - \sum_{A,B=1}^K \Lambda_{iA}~{\widetilde Q}^{-1}_{AB}~y_B\right) \leq w_i^-\label{Jm}
\end{eqnarray}
where ${\widetilde Q}^{-1}$ is the inverse of the $K\times K$ matrix ${\widetilde Q}$:
\begin{equation}
 {\widetilde Q}_{AB}\equiv \sum_{i\in{\widetilde J}} z_i~\Lambda_{iA}~\Lambda_{iB}
\end{equation}
Here the loadings matrix $\Lambda_{iA}$ must be such that ${\widetilde Q}$ is invertible.\footnote{\, This is the case if the columns of $\Lambda_{iA}$ are comprised of the first $K$ principal components of SCM $C_{ij}$ corresponding to its positive eigenvalues. However, as mentioned above, here we keep the loadings matrix general.} Also, note that $w_i$, $i\in {\widetilde J}$ given by (\ref{wv.bounds}) together with $w_i=w_i^+$, $i\in J^+$ and $w_i=w_i^-$, $i\in J^-$ satisfy (\ref{lin.const}), as they should.

{}Note that, for a given value of $\gamma$, (\ref{y}) solves for $y_A$ given $J^+$ and $J^-$. On the other hand, (\ref{Jp}) and (\ref{Jm}) determine $J^+$ and  $J^-$ in terms of $y_A$. The entire system is then solved iteratively, where at the initial iteration one takes ${\widetilde J}^{(0)}=J$, so that $J^{+(0)}$ and $J^{-(0)}$ are empty. However, we still need to fix $\gamma$. This is done via a {\em separate} iterative procedure, which we describe below.

{}Because we have two iterations, to guarantee (rapid) convergence, the $J^\pm$ iteration (that is, for a given value of $\gamma$) can be done as follows. Let ${\widehat w}_i^{(s)}$ be such that
\begin{eqnarray}
 &&\forall i\in J:~~~w_i^-\leq {\widehat w}^{(s)}_i\leq w_i^+\\
 &&\forall A\in \{1,\dots,K\}:~~~\sum_{i=1}^N {\widehat w}^{(s)}_i~\Lambda_{iA} = 0
\end{eqnarray}
At the $(s+1)$-th iteration, let $w^{(s+1)}_i$ be given by (\ref{wv.bounds}) for $i\in {\widetilde J}^{(s)}$, with $w^{(s+1)}_i=w^\pm_i$ for $i\in J^{\pm(s)}$. This solution satisfies (\ref{lin.const}), but may not satisfy the bounds. Let
\begin{eqnarray}
 &&q_i \equiv w^{(s+1)}_i - {\widehat w}^{(s)}_i\\
 &&h_i(t) \equiv {\widehat w}^{(s)}_i + t~q_i,~~~t\in[0,1]
\end{eqnarray}
Then
\begin{equation}\label{w-hat}
 {\widehat w}^{(s+1)}_i\equiv h_i(t_*) = {\widehat w}^{(s)}_i + t_*~q_i
\end{equation}
where $t_*$ is the maximal value of $t$ such that $h_i(t)$ satisfies the bounds. We have:
\begin{eqnarray}
 &&q_i > 0:~~~p_i \equiv \mbox{min}\left(w^{(s+1)}_i, ~w_i^+\right)\\
 &&q_i < 0:~~~p_i \equiv \mbox{max}\left(w^{(s+1)}_i, ~w_i^-\right)\\
 &&t_* = \mbox{min}\left({{p_i-{\widehat w}^{(s)}_i}\over q_i} ~\Big|~ q_i\neq 0,~i\in J\right)
\end{eqnarray}
Now, at each step, instead of (\ref{Jp}) and (\ref{Jm}), we can define $J^{\pm(s+1)}$ via
\begin{eqnarray}
 &&\forall i\in J^{+(s+1)}:~~~{\widehat w}^{(s+1)}_i = w_i^+\label{b+}\\
 &&\forall i\in J^{-(s+1)}:~~~{\widehat w}^{(s+1)}_i = w_i^-\label{b-}
\end{eqnarray}
where ${\widehat w}^{(s+1)}_i$ is computed iteratively as above and we can take ${\widehat w}_i^{(0)} \equiv 0$ at the initial iteration. Unlike (\ref{Jp}) and (\ref{Jm}), (\ref{b+}) and (\ref{b-}) add new elements to $J^\pm$ one (or a few) element(s) at each iteration.

{}The convergence criteria are given by
\begin{eqnarray}
 &&J^{+(s + 1)} = J^{+(s)}\\
 &&J^{-(s + 1)} = J^{-(s)}
\end{eqnarray}
These criteria are based on discrete quantities and are unaffected by computational (machine) precision effects. However, in practice the equalities in (\ref{b+}) and (\ref{b-}) are understood within some tolerance (or machine precision) -- see the R code in Appendix \ref{appA}. We will denote the value of ${\widehat w}^{(s+1)}_i$ at the final iteration (for a given value of $\gamma$, that is) via ${\widetilde w}_i$.

{}Finally, $\gamma$ is determined via another iterative procedure as follows (we use superscript $a$ for the $\gamma$ iterations to distinguish it from the superscript $s$ for the $J^\pm$ iterations):
\begin{equation}
 \gamma^{(a+1)} = {\gamma^{(a)}\over\sum_{i=1}^N \left|{\widetilde w}^{(a)}_i\right|}
\end{equation}
where ${\widetilde w}^{(a)}_i$ is computed as above for $\gamma=\gamma^{(a)}$. To achieve rapid convergence, the initial value $\gamma^{(0)}$ can be set as follows:
\begin{equation}
 \gamma^{(0)} = {1\over\sum_{i=1}^N z_i \left|\varepsilon_i\right|}
\end{equation}
where $\varepsilon_i$ are the residuals of the weighted regression (without bounds) given by (\ref{residuals}). The convergence criterion for the $\gamma$ iteration is given by
\begin{equation}
 \gamma^{(a+1)} = \gamma^{(a)}
\end{equation}
understood within some preset computational tolerance (or machine precision).

{}The R code for the above algorithm with some additional explanatory documentation is given in Appendix \ref{appA}. Note that this code is not written to be ``fancy" or optimized for speed or in any other way. Instead, its sole purpose is to {\em illustrate} the bounded regression algorithm as it is described above in a simple-to-understand fashion. Some legalese relating to this code is given in Appendix \ref{appB}.

\subsection{Application to Stock Portfolios}

{}Above we discussed the bounded regression algorithm in the context of computing weights for portfolios of alpha streams. However, the algorithm is quite general and -- with appropriate notational identifications -- can be applied to portfolios of stocks or other suitable instruments. In fact, it can also be applied outside of finance. Here, for the sake of definiteness, we will focus on stock portfolios, in fact, we will assume that they are dollar neutral, so both long and short positions are allowed.\footnote{\ Various generalizations are possible, some more straightforward than others.}

\subsubsection{Establishing Trades}\label{2.5.1}

{}Let us first discuss establishing trades, {\em i.e.}, we start from nil positions and establish a portfolio of $N$ stocks. Instead of alpha streams, our index $i\in\{1,\dots,N\} \equiv J$ now labels the stocks. We will denote the desired {\em dollar} (not share) holdings via $H_i$, and the total dollar investment (long plus short) via $I$:
\begin{equation}
 I\equiv\sum_{i=1}^N |H_i|
\end{equation}
Let $w_i\equiv H_i/I$. These are now our stock weights (analogous to the alpha weights). Then we have the familiar normalization condition
\begin{equation}
 \sum_{i=1}^N |w_i| = 1
\end{equation}
However, normally, one imposes bounds on $H_i$, not on $w_i$. For example, in the case of establishing trades one may wish to cap the positions such that: i) not more than a small percentile $\xi$ of the total dollar investment $I$ is allocated to any given stock -- this is a diversification constraint; and ii) only a small percentile ${\widetilde \xi}$ of ADDV (average daily dollar volume) $V_i$ is traded -- this is a liquidity constraint (see below). In this case we have the following bounds on the dollar holdings $H_i$:
\begin{eqnarray}
 &&H^-_i \leq H_i \leq H^+_i\\
 &&H^\pm_i = \pm\mbox{min}\left(\xi~I,~ {\widetilde \xi}~V_i\right)
\end{eqnarray}
In this case the upper and lower bounds are symmetrical. In some cases, such as for hard-to-borrow-stocks, we may have some $H^-_i = 0$. In other cases one may not wish to have a long position in some stocks. {\em Etc.} We will only assume that $H^-_i\leq 0$ and $H^+_i\geq 0$, in line with our discussion above for the bounds on the weights, which are then given by
\begin{equation}
 w^\pm_i \equiv H^\pm_i / I
\end{equation}
The final touch then is that instead of $\alpha_i$ one uses some expected returns $E_i$ in the case of stocks. The rest goes through exactly as above for a suitably chosen $\Lambda_{iA}$.

\subsubsection{Rebalancing Trades}

{}With rebalancing trades, we have the current dollar holdings $H^*_i$ and the desired dollar holdings $H_i$. In this case, one may wish to cap the positions such that: i) not more than a small percentile $\xi$ of the total dollar investment $I$ is allocated to any given stock -- this is the same diversification constraint as above; ii) only a small percentile ${\widetilde \xi}$ of ADDV $V_i$ is traded -- this the same liquidity constraint as above; and iii) not more than a small percentile $\xi^\prime$ of ADDV $V_i$ is allocated to any given stock -- this is another liquidity constraint stemming from the consideration that, if the portfolio must be liquidated swiftly ({\em e.g.}, due to an unforeseen event), to mitigate liquidation costs, the positions are capped based on liquidity. Here $\xi^\prime$ typically can be several times larger than ${\widetilde \xi}$ -- the portfolio can be built up in stages as long at each stage the bounds are satisfied. The bounds on $H_i$ now read:
\begin{eqnarray}
 && |H_i| \leq \mbox{min}\left(\xi~ I,~\xi^\prime~V_i\right)\\
 && |H_i - H_i^*| \leq {\widetilde \xi}~V_i
\end{eqnarray}
It is more convenient to rewrite these bounds in terms of the traded dollar amounts $D_i \equiv H_i - H_i^*$:
\begin{eqnarray}
 && D_i^- \leq D_i \leq D_i^+\\
 && D_i^+ = \mbox{min}\left(\mbox{min}\left(\xi~ I,~\xi^\prime~V_i\right) - H_i^*,~{\widetilde \xi}~V_i\right)\geq 0\\
 && D_i^- = \mbox{max}\left(-\mbox{min}\left(\xi~ I,~\xi^\prime~V_i\right) - H_i^*,~-{\widetilde \xi}~V_i\right)\leq 0
\end{eqnarray}
and we are assuming that $|H_i^*| \leq \mbox{min}\left(\xi~ I,~\xi^\prime~V_i\right)$. Furthermore, we will assume that $H_i^*$ itself satisfies (\ref{lin.const}):
\begin{equation}
 \forall A\in\{1,\dots,K\}:~~~\sum_{i=1}^N H_i^*~\Lambda_{iA} = 0
\end{equation}
Then the bounded regression algorithm can be straightforwardly applied to the weights $w_i$ and $x_i$ defined as follows:
\begin{eqnarray}
 &&w_i \equiv H_i/I\\
 &&x_i \equiv D_i/I
\end{eqnarray}
In the $J^\pm$ iteration we now use $x_i$ instead of $w_i$, while in the $\gamma$ iteration we still use $w_i$. Then the rest of the algorithm goes through unchanged. Let us note, however, that the source code given in Appendix \ref{appA} is written with alpha weights in mind, so while it can be adapted to the case of stock portfolios in the case of establishing trades, straightforward modifications are required to accommodate rebalancing trades.

\subsubsection{Examples: Intraday Mean-Reversion Alphas}\label{2.5.3}

{}To illustrate the use of the algorithm, we have employed it to construct portfolios for intraday mean-reversion alphas with the loadings matrix $\Lambda_{iA}$ in the following 5 incarnations: i) intercept only (so $K=1$); ii) BICS (Bloomberg Industry Classification System) sectors; iii) BICS industries; iv) BICS sub-industries; and v) the 4 style factors prc, mom, hlv and vol of \cite{4FM} plus BICS sub-industries. The regression weights are the inverse sample variances: $z_i = 1/C_{ii}$ (see below). In the cases ii)-v) above the intercept is subsumed in the loadings matrix $\Lambda_{iA}$. Indeed, we have $\sum_{A\in G} \Lambda_{iA} \equiv 1$, where $G$ is the set of columns of $\Lambda_{iA}$ corresponding to sectors in the case ii), industries in the case iii), and sub-industries in the cases iv) and v). Consequently, the resultant portfolios are automatically dollar neutral.

{}The portfolio construction and backtesting are identical to those in \cite{RussDoll}, where more detailed discussion can be found, so to save space, here we will only give a brief summary. The portfolios are assumed to be established at the open and liquidated at the close on the same day, so they are purely intraday and the algorithm of Section \ref{2.5.1} for the establishing trades applies. The expected returns $E_i$ for each date are taken to be $E_i = -R_i$, where $R_i \equiv \ln\left(P_i^{open}/P_i^{close}\right)$, and for each date $P_i^{open}$ is today's open, while $P_i^{close}$ is yesterday's close adjusted for splits and dividends if the ex-date is today. So, these are intraday mean-reversion alphas.

{}The universe is top 2000 by ADDV $V_i$, where ADDV is computed based on 21-trading-day rolling periods. However, the universe is not rebalanced daily, but also every 21 trading days (see \cite{RussDoll} for details). The sample variances $C_{ii}$ are computed based on the same 21-trading-day rolling periods, and are not applied daily, but also every 21 trading days, same as the universe rebalancing (see \cite{RussDoll} for details). We run our simulations over a period of 5 years (more precisely, $252 \times 5$ trading days going back from 9/5/2014, inclusive). The annualized return-on-capital (ROC) is computed as average daily P\&L divided by the total (long plus short) intraday investment level $I$ (with no leverage) and multiplied by 252. The annualized Sharpe Ratio (SR) is computed as the daily Sharpe ratio multiplied by $\sqrt{252}$. Cents-per-share (CPS) is computed as the total P\&L divided by the total shares traded. On each day the total (establishing plus liquidating) shares traded for each stock are given by $Q_i = 2|H_i|/P_i^{open}$ (see \cite{RussDoll} for details).

{}For comparison purposes, the results for regressions without bounds are given in Table \ref{table1}. The results for the bounded regressions, with the bounds on the desired holdings set as
\begin{equation}
 |H_i| \leq 0.01~V_i
\end{equation}
so not more than 1\% of each stock's ADDV is bought or sold, are given in Table \ref{table2} and the corresponding P\&Ls are plotted in Figure 1. Thus, as expected, adding the liquidity bounds has the diversification effect on the portfolios, so the Sharpe ratios are substantially improved -- as usual, at the expense of (slightly) lowering paper ROC and CPS. Note that, even with tight liquidity bounds, the 4 style factors prc, mom, hlv and vol of \cite{4FM} add value, further validating the 4-factor model of \cite{4FM}.

\section{Concluding Remarks}\label{sec3}

{}One -- but not the only -- way to think about bounded regression is as an alternative to optimization with bounds when the latter is not attainable. In fact, as mentioned above, bounded regression is a zero specific risk limit of optimization with bounds in the context of a factor model. So, when a factor model is not available, {\em e.g.}, in the context of alpha streams, bounded regression can be used in lieu of optimization.

{}In this regard, one can further augment the bounded regression algorithm we discussed above by including linear transaction costs, as in \cite{AlphaOpt}. A systematic approach is to start with optimization with bounds and linear transaction costs in the context of a factor model as in \cite{MeanRev} and take a zero specific risk limit. Non-linear transaction costs (impact) in the context of alpha weights can be treated using the approximation discussed in \cite{AlphaOpt} using the spectral model of turnover reduction \cite{SpMod}.

\appendix

\section{The R Code}\label{appA}

{}Below we give R (R Package for Statistical Computing, http://www.r-project.org) source code for the bounded regression algorithm we discuss in the main text. The entry function is {\tt{\small calc.bounded.lm()}}, which runs the $\gamma$ iteration loop and calls the function {\tt{\small bounded.lm()}}, which runs the $J^\pm$ iteration loop. The {\tt{\small args()}} of {\tt{\small calc.bounded.lm()}} are: {\tt{\small ret}}, which is the $N$-vector of alphas $\alpha_i$ (or, more generally, some other returns); {\tt{\small load}}, which is the $N\times K$ loadings matrix $\Lambda_{iA}$; {\tt{\small weights}}, which is the $N$-vector of the regression weights $z_i$; {\tt{\small upper}}, which is the $N$-vector of the upper bounds $w^+_i$; {\tt{\small lower}}, which is the $N$-vector of the lower bounds $w_i^-$; and {\tt{\small prec}}, which is the desired precision with which the output weights $w_i$, the $N$-vector of which {\tt{\small calc.bounded.lm()}} returns, must satisfy the normalization condition (\ref{w.norm}). Internally, {\tt{\small bounded.lm()}} calls the function {\tt{\small calc.bounds()}}, which computes ${\widehat w}^{(s+1)}_i$ in (\ref{w-hat}) at each iteration. The code is straightforwardly self-explanatory. {\tt{\small Jp, Jm}} in {\tt{\small bounded.lm()}} correspond to $J^\pm$. One subtlety is that, when restricting $\Lambda_{iA}$ to ${\widetilde J}\subset J$, in the case of binary industry classification ({\em e.g.}, when $\Lambda_{iA}$ corresponds to BICS sub-industries, which can be small), the so-restricted $\Lambda_{iA}$ may have null columns, which must be omitted and the code below does just that. For non-binary cases, one may wish to augment the code to ensure that the matrix {\tt{\small Q <- t(load[Jt, take]) \%*\% w.load[Jt, take]}} is nonsingular (and if it is, then remove the culprit columns in $\Lambda_{iA}$ or otherwise modify the latter); however, for non-binary $\Lambda_{iA}$ and generic regression weights this should not occur except for special, non-generic cases.\\
\\
{\tt{\small
\noindent calc.bounded.lm <- function(ret, load, weights, upper, lower, prec = 1e-5)\\
\{\\
\indent	reg <- lm(ret $\sim$ -1 + load, weights = weights)\\
\indent	x <- weights * residuals(reg)\\
\indent	ret <- ret / sum(abs(x))\\
\\
\indent	repeat\{\\
\indent\indent		x <- bounded.lm(ret, load, weights, upper, lower)\\
\indent\indent		if(abs(sum(abs(x)) - 1) < prec)\\
\indent\indent\indent			break\\

\indent\indent		ret <- ret / sum(abs(x))\\
\indent	\}\\
\\
\indent	return(x)\\
\}\\
\\
bounded.lm <- function(ret, load, weights, upper, lower, tol = 1e-6)\\
\{\\
\indent	calc.bounds <- function(z, x)\\
\indent	\{\\
\indent\indent		q <- x - z\\
\indent\indent		p <- rep(NA, length(x))\\
\indent\indent		pp <- pmin(x, upper)\\
\indent\indent		pm <- pmax(x, lower)\\
\indent\indent		p[q > 0] <- pp[q > 0]\\
\indent\indent		p[q < 0] <- pm[q < 0]\\
\indent\indent		t <- (p - z)/q\\
\indent\indent		t <- min(t, na.rm = T)\\
\indent\indent		z <- z + t * q\\
\indent\indent		return(z)\\
\indent	\}\\
\\
\indent	if(!is.matrix(load))\\
\indent\indent		load <- matrix(load, length(load), 1)\\
\\
\indent	n <- nrow(load)\\
\indent	k <- ncol(load)\\
\\
\indent	ret <- matrix(ret, n, 1)\\
\indent	upper <- matrix(upper, n, 1)\\
\indent	lower <- matrix(lower, n, 1)\\
\indent	z <- diag(weights)\\
\indent	w.load <- z \%*\% load\\
\indent	w.ret <- z \%*\% ret\\
\\
\indent	J <- rep(T, n)\\
\indent	Jp <- rep(F, n)\\
\indent	Jm <- rep(F, n)\\
\indent	z <- rep(0, n)\\
\\
\indent	repeat\{\\
\indent\indent		Jt <- J \& !Jp \& !Jm\\
\indent\indent		y <- t(w.load[Jt, ]) \%*\% ret[Jt, ]\\
\indent\indent		if(sum(Jp) > 1)\\
\indent\indent\indent			y <- y + t(load[Jp, ]) \%*\% upper[Jp, ]\\
\indent\indent		else if(sum(Jp) == 1)\\
\indent\indent\indent			y <- y + upper[Jp, ] * matrix(load[Jp, ], k, 1)\\
\indent\indent		if(sum(Jm) > 1)\\
\indent\indent\indent		y <- y + t(load[Jm, ]) \%*\% lower[Jm, ]\\
\indent\indent		else if(sum(Jm) == 1)\\
\indent\indent\indent			y <- y + lower[Jm, ] * matrix(load[Jm, ], k, 1)\\
\indent\indent		if(k > 1)\\
\indent\indent\indent			take <- colSums(abs(load[Jt, ])) > 0\\
\indent\indent		else\\
\indent\indent\indent			take <- T\\
\indent\indent		Q <- t(load[Jt, take]) \%*\% w.load[Jt, take]\\
\indent\indent		Q <- solve(Q)\\
\indent\indent		v <- Q \%*\% y[take]\\
\\
\indent\indent		xJp <- Jp\\
\indent\indent		xJm <- Jm\\
\\	
\indent\indent		x <- w.ret - w.load[, take] \%*\% v\\
\indent\indent		x[Jp, ] <- upper[Jp, ]\\
\indent\indent		x[Jm, ] <- lower[Jm, ]\\
\\
\indent\indent		z <- calc.bounds(z, x)\\
\indent\indent		Jp <- abs(z - upper) < tol\\
\indent\indent		Jm <- abs(z - lower) < tol\\
\\		
\indent\indent		if(all(Jp == xJp) \& all(Jm == xJm))\\
\indent\indent\indent			break\\
\indent	\}\\
\\
\indent	return(z)\\		
\}
}}

\section{DISCLAIMERS}\label{appB}

{}Wherever the context so requires, the masculine gender includes the feminine and/or neuter, and the singular form includes the plural and {\em vice versa}. The author of this paper (``Author") and his affiliates including without limitation Quantigic$^\circledR$ Solutions LLC (``Author's Affiliates" or ``his Affiliates") make no implied or express warranties or any other representations whatsoever, including without limitation implied warranties of merchantability and fitness for a particular purpose, in connection with or with regard to the content of this paper including without limitation any code or algorithms contained herein (``Content").

{}The reader may use the Content solely at his/her/its own risk and the reader shall have no claims whatsoever against the Author or his Affiliates and the Author and his Affiliates shall have no liability whatsoever to the reader or any third party whatsoever for any loss, expense, opportunity cost, damages or any other adverse effects whatsoever relating to or arising from the use of the Content by the reader including without any limitation whatsoever: any direct, indirect, incidental, special, consequential or any other damages incurred by the reader, however caused and under any theory of liability; any loss of profit (whether incurred directly or indirectly), any loss of goodwill or reputation, any loss of data suffered, cost of procurement of substitute goods or services, or any other tangible or intangible loss; any reliance placed by the reader on the completeness, accuracy or existence of the Content or any other effect of using the Content; and any and all other adversities or negative effects the reader might encounter in using the Content irrespective of whether the Author or his Affiliates is or are or should have been aware of such adversities or negative effects.

{}The R code included in Appendix \ref{appA} hereof is part of the copyrighted R code of Quantigic$^\circledR$ Solutions LLC and is provided herein with the express permission of Quantigic$^\circledR$ Solutions LLC. The copyright owner retains all rights, title and interest in and to its copyrighted source code included in Appendix \ref{appA} hereof and any and all copyrights therefor.

\begin{table}[ht]
\caption{Simulation results for the 5 alphas via regression without bounds discussed in Section \ref{2.5.3}.} 
\begin{tabular}{l l l l} 
\\
\hline\hline 
Alpha & ROC & SR & CPS\\[0.5ex] 
\hline 
Regression: Intercept only & 33.59\% & 5.59 & 1.38\\
Regression: BICS Sectors & 39.28\% & 7.05 & 1.61\\
Regression: BICS Industries & 42.66\% & 8.19 & 1.75\\
Regression: BICS Sub-industries & 45.25\% & 9.22 & 1.84\\
Regression: 4 Style Factors plus BICS Sub-industries & 46.60\% & 9.85 & 1.90\\[1ex] 
\hline 
\end{tabular}
\label{table1} 
\end{table}

\begin{table}[ht]
\caption{Simulation results for the 5 alphas via bounded regression discussed in Section \ref{2.5.3}.} 
\begin{tabular}{l l l l} 
\\
\hline\hline 
Alpha & ROC & SR & CPS\\[0.5ex] 
\hline 
Regression: Intercept only & 29.66\% & 7.36 & 1.25\\
Regression: BICS Sectors & 35.32\% & 9.89 & 1.48\\
Regression: BICS Industries & 39.25\% & 12.00 & 1.65\\
Regression: BICS Sub-industries & 42.23\% & 14.13 & 1.75\\
Regression: 4 Style Factors plus BICS Sub-industries & 43.70\% & 15.54 & 1.82\\[1ex] 
\hline 
\end{tabular}
\label{table2} 
\end{table}

\newpage
\begin{figure}[ht]
\centerline{\epsfxsize 4.truein \epsfysize 4.truein\epsfbox{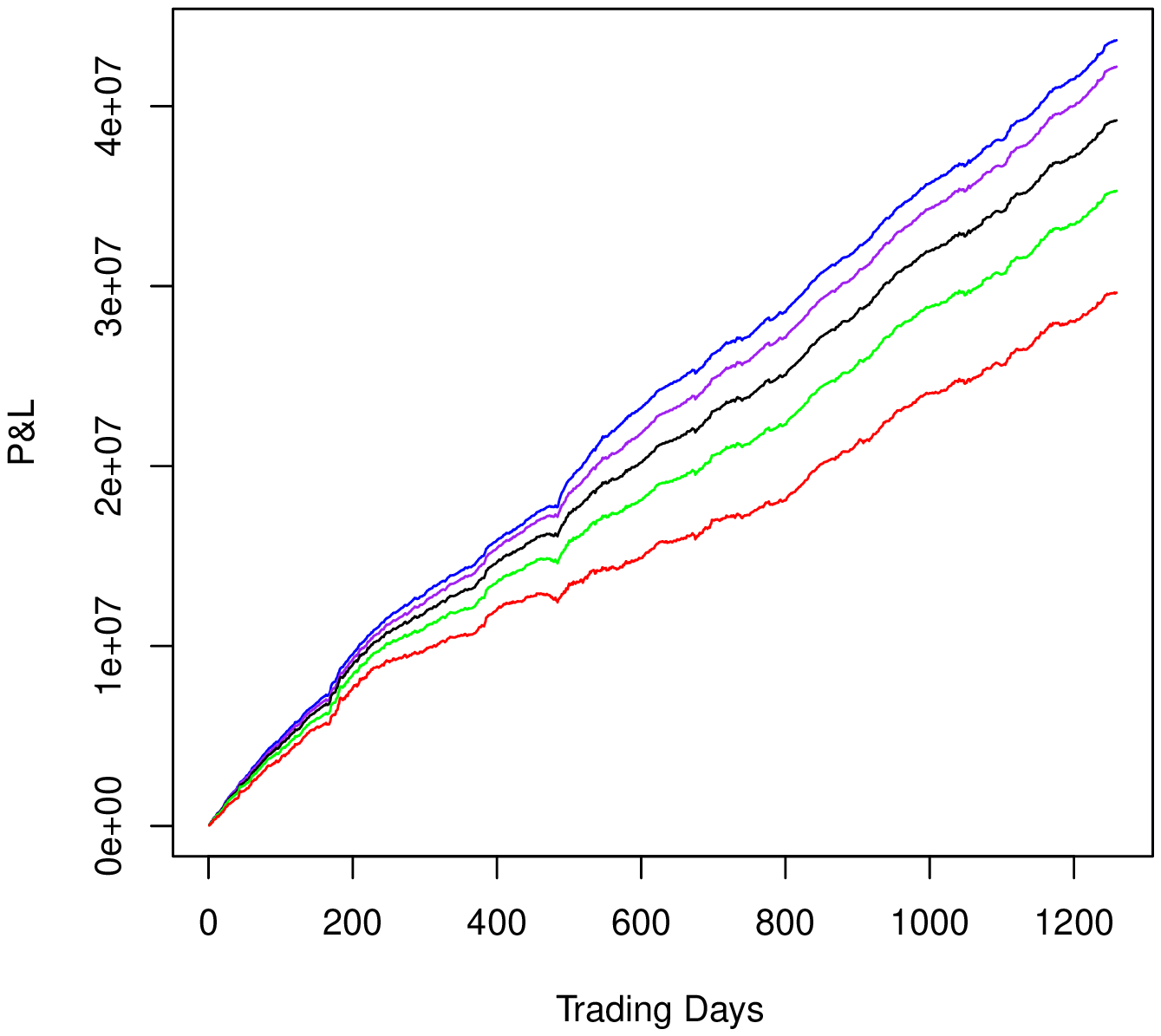}}
\noindent{\small {Figure 1. P\&L graphs for the intraday alphas discussed in Section \ref{2.5.3}, with a summary in Table \ref{table2}. Bottom-to-top-performing: i) Regression over intercept only, ii) regression over BICS sectors, iii) regression over BICS industries, iv) regression over BICS sub-industries, and v) regression over 4 style factors prc, mom, hlv and vol of \cite{4FM} plus BICS sub-industries. The investment level is \$10M long plus \$10M short.}}

\end{figure}

\end{document}